\documentclass[12pt,onecolumn,final]{IEEEtran}                  
\usepackage{graphicx}
\usepackage{amsmath}
\usepackage{amsfonts}
\usepackage{amssymb}
\usepackage{color}

\newcommand{\blue}{}

\def\blfootnote{\xdef\@thefnmark{}\@footnotetext} 
\long\def\symbolfootnote[#1]#2{\begingroup
\def\thefootnote{\fnsymbol{footnote}}\footnote[#1]{#2}\endgroup}
\begin{document}
\author{Amir Rosenthal, Alex Linden, and Moshe Horowitz%
\thanks{Accepted for publication in JOSA A\newline
The authors are with the Technion, Israel Institute of Technology, Haifa, Israel, 32000,  eeamir@tx.technion.ac.il, alinden@ee.technion.ac.il, horowitz@ee.technion.ac.il}}
\title{Multi-rate asynchronous sampling of sparse multi-band signals}
\date{}
\maketitle

\begin{abstract}
\blue{Because optical systems have huge bandwidth and are capable of generating low noise short pulses
they are ideal for undersampling
multi-band signals that are located within a very broad frequency range.
In this paper we
propose a new scheme for reconstructing
multi-band signals that occupy a small part of a given broad frequency range
under the constraint of a small number of sampling channels.} The
scheme, which we call multi-rate sampling (MRS), entails gathering
samples at several different rates whose sum is significantly lower
than the Nyquist sampling rate. The number of channels does not
depend on any characteristics of a signal.
In order to be
implemented with simplified hardware, the
reconstruction method does not rely on the synchronization between
different sampling channels. Also, because the method does not solve a system of
linear equations, it avoids one source of lack of robustness of
previously published undersampling schemes. Our simulations indicate
that our MRS scheme is robust both to different signal types and to
relatively high noise levels. The scheme can be implemented easily with
optical sampling systems.
\end{abstract}


\section{Introduction}
A multi-band signal is one whose energy in the frequency domain is
contained in the finite union of closed intervals. A sparse signal
is a signal that occupies only a small portion of a given frequency region.
In many applications of radars and communications systems it
is desirable to reconstruct a multi-band sparse signal from its
samples. When the signal bands are centered at frequencies that are high
compared to their widths, it is not cost effective and
often it is not feasible to sample at the Nyquist rate $F_{\textrm{nyq}}$; the rate that for
a real signal is equal to twice the maximum frequency of the given region in which the signal spectrum is located.
It is therefore desirable to reconstruct the signal by undersampling;
that is to say, from samples taken at rates significantly lower than the Nyquist
rate. Sampling at any constant rate that is lower than the Nyquist
rate results in down-conversion of all signal bands to a low
frequency region called a baseband. This creates two problems in the
reconstruction of the signal. The first is a loss of knowledge of
the actual signal frequencies. The second is the possibility of aliasing;
i.e. spectrum at different frequencies being down-converted to
the same frequency in the baseband.


Optical systems are capable of very high performance undersampling
\cite{Avi}. They can handle signals whose carrier frequency can be very
high, on the order of 40 GHz, and signals with a dynamic range as high as 70 dB.
The size, the weight, and the power consumption
of optical systems make them ideal for undersampling. The
simultaneous sampling of a signal at different time offsets or at different rates can
be performed efficiently by using techniques based on
wavelength-division multiplexing (WDM) that are used in optical
communication systems.

There is a vast literature on reconstructing signals from
undersampled data.  Landau proved that, regardless of the sampling
scheme, it is impossible to reconstruct a signal of spectral measure
$\lambda$ with samples taken at an average rate less than
$\lambda$~\cite{Landau}. This rate $\lambda$ is commonly referred to
as the Landau rate.  Much work has been done to develop schemes that
can reconstruct signals at sampling rates close to the Landau rate.
Most are a form of a periodic nonuniform sampling (PNS) scheme
\cite{Kohlenberg}-\cite{Herley}. Such a scheme was introduced by
Kohlenberg~\cite{Kohlenberg} who applied it to a single-band signal
whose carrier frequency is known a priori. The PNS scheme was later
extended to reconstruct multi-band signals with carrier frequencies
that are known a priori \cite{Venkantaramani_Bresler}, \cite{Lin}.

In a PNS scheme $m$ low-rate cosets are chosen out of $L$ cosets of
samples obtained from time-uniformly distributed samples taken at a
rate $F$ where $F$ is greater or equal to the Nyquist rate $F_{\textrm{nyq}}$
\cite{Venkantaramani_Bresler}. Consequently, the sampling rate of
each sampling channel is $L$ times lower than $F$ and the overall
sampling rate is $L/m$ times lower than $F$. The samples obtained
from the sampling channels are offset by an \blue{integral multiple of a constant time increment, $1/F$.} This  \blue{sampling scheme}
may resolve aliasing. In a PNS scheme the signal is reconstructed by solving a system
of linear equations~\cite{Venkantaramani_Bresler}. PNS schemes can often achieve
perfect reconstructions from samples taken at a rate that approaches
the Landau rate under the assumption that the carrier frequencies
are known a priori. However, in order
to attain a perfect reconstruction, the number of sampling channels
must be sufficiently high such that the equations have a unique
solution \cite{Venkantaramani_Bresler}.

When the carrier frequencies of the signals are not known a priori, in a PNS scheme, a perfect
reconstruction requires the sampling rate to exceed twice
the Landau rate \cite{DO,Moshiko}. In addition, in a PNS scheme the
number of sampling channels must be sufficiently high
\cite{Moshiko}. Under these two conditions, a solution to the
set of equations in PNS scheme may be obtained assuming that the sampled
signal is sparse \cite{Moshiko}. When a PNS scheme is applied to an
$N$-band real signal ($N$ bands in the interval $[0,F_{\textrm{nyq}}/2]$), at
least $4N$ channels are required for a perfect reconstruction
\cite{DO,Moshiko}. A method for obtaining a perfect
reconstruction has been demonstrated only with the number of channels
equal to $8N$ \cite{Moshiko}. Even when the requirement of perfect
reconstruction is relaxed, the number of channels required to obtain
an acceptably small error in the reconstructed signal may be prohibitively large. Furthermore,
the implementation of the schemes to attain the minimum
sampling rate relies heavily on assumed values of the
widths of the sample bands and the number of bands of the signal \cite{Moshiko}. In the case that
the bands of the signal have different widths, a PNS scheme for
obtaining the minimum sampling rate has not been demonstrated.

Other important drawbacks of PNS schemes stem from the fact that the
systems of equations to be solved are poorly conditioned
\cite{Feng_Bresler}. Thus, the schemes are sensitive to the bit
number of A/D conversion. They are also sensitive to any noise
present in a signal and to the spectrum of the signal at any
frequencies outside of strictly defined bands. Moreover, the use of
undersampling significantly increases the noise in each sampling
channel since the noise in the entire sampled spectrum is
downconverted to low frequencies. Therefore, the dynamic range of
of the overall system is limited. The noise may be reduced by increasing the
sampling rate in each channel. However, since the number of channels
needed for a perfect reconstruction is determined only by the number of signal bands,
the overall sampling rate dramatically increases.
Another important
drawback of PNS scheme is the requirement of a very low time jitter
between the samplings in the different channels.

In this paper we propose a different scheme for reconstructing
sparse multi-band signals. The scheme, which we call multi-rate
sampling (MRS), entails gathering samples at $P$ different rates.
The number $P$ is small (three in our simulations) and does not
depend on any characteristics of a signal. Our approach is not
intended to obtain the minimum sampling rate. Rather, it is
intended to reconstruct signals accurately with a very high
probability at an overall sampling rate that is significantly lower than the
Nyquist rate under the constraint of a small number of channels.

The success of our MRS scheme relies on the assumption that sampled
signals are sparse. For a typical sparse signal, most of the sampled
spectrum is unaliased in at least one of the $P$ channels. This is
in contrast to the situation that prevails with PNS schemes. In PNS
schemes, because all channels are sampled at the same frequency, an
alias in one channel is equivalent to an alias in all channels.

In our MRS scheme, the sampling rate of each channel is chosen to be
approximately equal to the maximum sampling rate allowed by cost and
technology. Consequently, in most applications, the sampling rate is
significantly higher than twice the maximum width of the signal
bands as usually assumed in PNS schemes.

Sampling at higher rates has a fundamental advantage if
signals are contaminated by noise. The spectrum evaluated at a
baseband frequency $f_b$ in a channel sampling at a rate $F$ is the
sum of the spectrum of the original signal at all frequencies
$f_b+mF$ that are located in the system bandwidth, where $m$ 
ranges over all integers. Thus, the larger the value of $F$, the fewer terms
contribute to this sum.  As a result, sampling at a higher rate
increases the signal to noise ratio in the base-band region.

\blue{To} simplify the hardware needed for the sampling, our
reconstruction method was developed to not require synchronization between
different sampling channels. Therefore, our method enables a significant reduction in
the complexity of the hardware. Moreover,
unsynchronized sampling relaxes the stringent requirement in PNS
schemes of a very small timing jitter in the sampling time of the
channels. We also do not need to solve a linear set of
equations. This eliminates one source of lack of robustness of PNS
schemes. Our simulations indicate that MRS schemes are robust both
to different signal types and to relatively high noise. The ability
of our MRS scheme to reconstruct parts of the signal spectrum that alias when sampled
at all $P$ sampling rates can be enhanced by using more complicated
hardware that synchronizes all of the sampling channels.

The paper is organized as follows.  In section 2 we present some
general mathematical background. In section 3 we describe the
algorithm.  In section 4 we give some considerations regarding our
algorithm complexity. In section 5 we present results of computer
simulations.

\section{Mathematical Background and Notation}

A multi-band signal is one whose energy in the frequency domain is
contained in a finite union of closed intervals \blue{$\bigcup_{n=1}^N[a_i,b_i]$}. \blue{A multi-band signal $x(t)$ is said to be sparse in the interval $[F_{\textrm{min}},F_{\textrm{max}}] $ if the Lebesgue
measure of its spectral support $\lambda(x)=\sum_{n=1}^N (b_n-a_n)$
satisfies $\lambda \ll F_{\textrm{max}}-F_{\textrm{min}}$.}

The signals we consider are sparse multi-band with spectral measure
$\lambda$. We use the \blue{following form of the Fourier transform of a signal $x(t)$:
\begin{equation}\label{ft def}
X(f)=\int_{-\infty}^{\infty} x(t)\exp(-2\pi ift).
\end{equation}
}
If the signal $x(t)$ is real (as is every physical
signal), then its spectrum $X$ satisfies $X(f)=\overline{X}(-f)$
where $\overline{a+bi}=a-bi$ and $a$ and $b$ are real numbers. Thus,
a real multi-band signal $x(t)$ has fourier transform $X(f)$ which,
when decomposed into its support intervals, can be represented by
\begin{eqnarray}
X(f)=\sum_{n=1}^{N}
\left[S_n(f)+\overline{S}_n(-f)\right],\label{multi_band}
\end{eqnarray}
where $S_n(f)\neq0$ only for $f\in[a_n,b_n]$ (where $b_n>a_n\geq
0$), and $[a_n,b_n]\bigcap [a_m,b_m] = \phi $ for all $n\neq m$. 

We assume that $F_{\textrm{nyq}}$ is
known a priori. That is to say, we assume that each $b_n$ for a real signal is at most
some known value $F_{\textrm{nyq}}/2$.
Sampling a signal $x(t)$ at a uniform rate $F^i$ produces a sampled
signal \blue{
\begin{equation}\label{ft}
x^i(t)=x(t+\Delta^i)\sum_{n=-\infty}^\infty\delta\left(t-\frac{n}{F^i}\right),
\end{equation}
}
where $\Delta^i$ is a time offset between the clock of the sampling
system and a hypothetical clock that defines an absolute time for
the signal. \blue{Because we are assuming a lack of synchronization between more than one sampling channel, we assume that the time offsets $\Delta^i$ are unknown. Reconstructing the amplitude of the signal spectrum with our scheme does not require knowledge of the time offsets. Only in reconstructing the phase of the signal in the frequency domain, do we need in some cases to extract the differences between time offsets. }

\begin{figure}[htb]
\hspace{4.5 cm} \includegraphics[angle=-90, width=14cm]{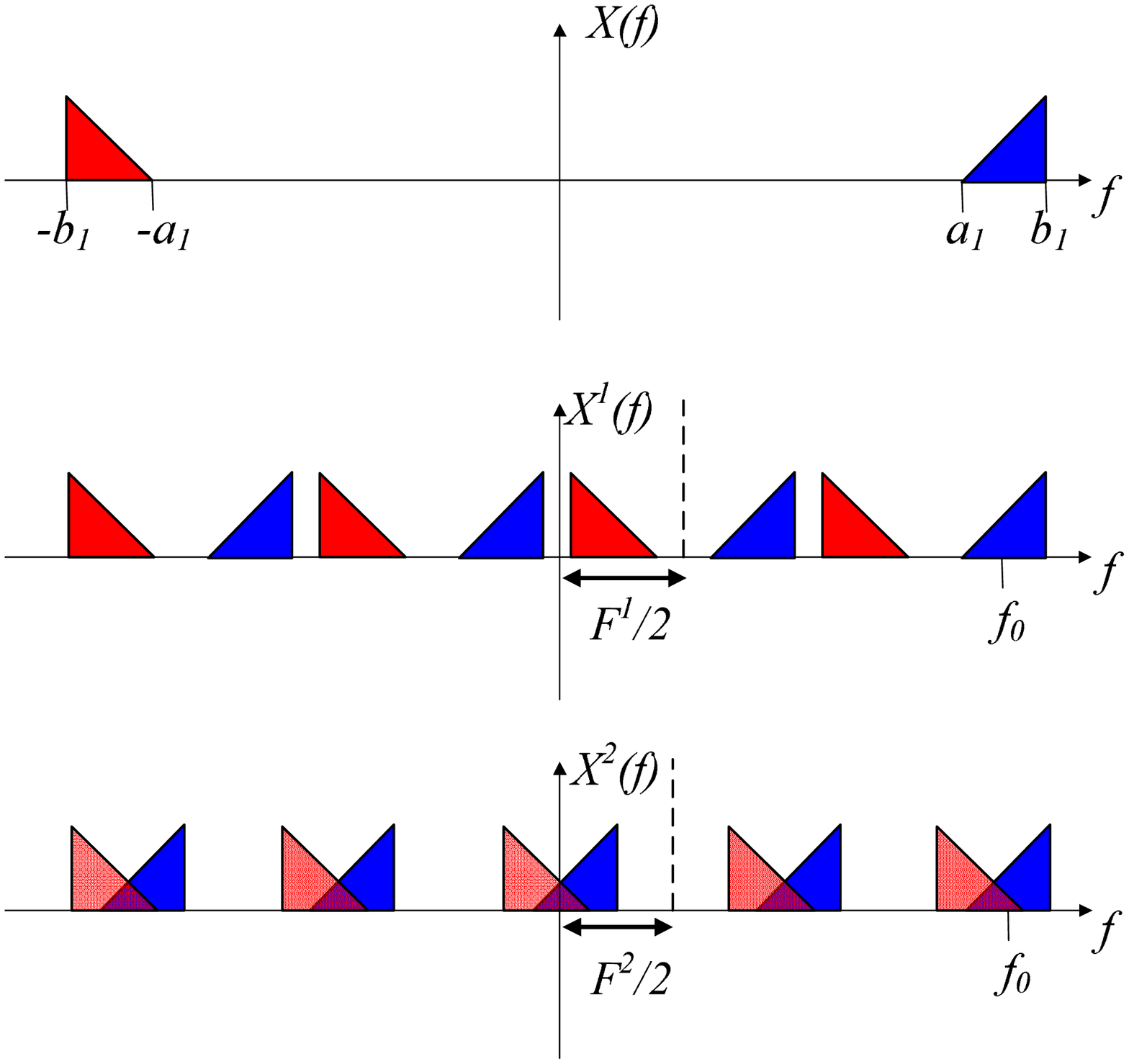}
\caption{\label{illustration}
Illustration of  the spectrum of  a sparse one-band real
signal (a), and the spectrum of its samples which are obtained for
the sampling rates  $F^1$ (b) and  $F^2$ (c). At $f_0$, the signal
is unaliased at the sampling rate $F^1$, but is aliased at the
sampling rate $F^2$.}
\end{figure}

The Fourier transform of a sampled signal $x^i(t)$, $X^i(f)$, is given by
\begin{equation}\label{baseband sum}
X^i(f)=F^i\sum_{n=-\infty}^\infty X(f+nF^i) \exp[2\pi
i(f+nF^i)\Delta^i].
\end{equation}
The connection between the spectrum of a sparse signal $X(f)$ and
the spectrum of its sampled signal $X^i(f)$ is illustrated in
Fig.~\ref{illustration}.

One immediate consequence of Eq.~\ref{baseband sum} is that\blue{, up to a phase factor that does not depend on the signal, $\exp[2\pi
i(f+nF^i)\Delta^i]$}, $X^i(f)$
is periodic of period $F^i$.  It is also clear that, for a real
signal $x(t)$, $\overline{X^i}(-f)=X^i(f)$. Thus, all of the
information \blue{about} \blue{$|X^i(f)|$} is contained in the interval $[0,F^i/2]$. Beside a linear chirp caused by the time offset $\Delta^i$ all the information about the phase of $X^i(f)$ is also contained in the interval $[0,F^i/2]$. We
shall refer to this interval $[0,F^i/2]$ as the $i$th baseband. The
down-conversion of a frequency $f\in[0,F_{\textrm{nyq}}/2]$ to this baseband
is represented by the down-conversion function
$D^i:[0,F_{\textrm{nyq}}/2]\longrightarrow [0,F^i/2]$:\blue{
\begin{equation}\label{downconvert}
D^i(f)=\min [ f \mod F^i,(F^i-f) \mod F^i ].
\end{equation}
}
In the case of band-limited signal $X(f)$, for a given frequency
$f$, all but a finite number of terms in the infinite sum on the
right side of Eq.~\ref{baseband sum} vanish. If the number of
non-vanishing terms is greater than one for a given sampling rate
$F^i$, then the signal is said to be aliased at $f$ when sampled at
the rate $F^i$. If at a frequency $f$ only a single term in the sum
is not equal to zero, the signal $X(f)$ is said to be unaliased at a
sampling rate $F^i$. Illustration of aliasing can be seen in
Fig.~\ref{illustration}(c). In the case of sparse signals, $x(t)$ is
unaliased at considerable part of it spectral support. The success
of an MRS scheme lies in the fact that whereas a signal may be
aliased at a frequency $f$ when sampled at a rate $F^i$, the same
signal may be unaliased at the same frequency $f$ when sampled at a
different rate $F^j$.

Each support interval $[a,b]$  ($b>a\geq 0$) of the multi-band
signal will be referred to as an \textit{originating band}.
According to Eq.~\ref{baseband sum}, sampling at the rate $F^i$
down-converts each originating band $[a,b]$
 to a single band in the baseband
$[\alpha^i,\beta^i]$.  We shall refer to the interval
$[\alpha^i,\beta^i]$ as a \textit{down-converted band}.

It is apparent that when a single down-converted band
$[\alpha^i,\beta^i]$ is given, it is in general not possible to
identify its corresponding originating band. However, it follows
easily from Eq.~\ref{baseband sum} that the corresponding
originating band must reside within the set of bands defined by
\begin{equation}\label{upconvert}
\left\{\left(\bigcup_{m=-\infty}^\infty\left[\alpha^i+mF^i,\beta^i+mF^i\right]\right)\bigcup\left(\bigcup_{m=
-\infty}^\infty\left[-\beta^i+mF^i,-\alpha^i+mF^i\right]\right)\right\}\bigcap\left[0,F_{\textrm{nyq}}/2\right],
\end{equation}
where $m$ is an integer. The set in Eq.~\ref{upconvert} can
be represented as a finite number of disjointed closed intervals,
which we denote by $[a_n^i,b_n^i]$. We shall refer to each of these
intervals as an \textit{up-converted band}. For clarity, we denote
all down-converted intervals with greek letters superscripted by the
sampling frequency and denote all up-converted intervals with latin
letters.

In general, the number of possible originating bands is reduced by
sampling at more than one rate. For each sampling rate rate $F^i$,
an originating band $[a,b]$ must reside within the union of the
upconverted bands: $[a,b]\in \cup_n [a_n^i,b_n^i]$. Since the union
of upconverted bands is different for each sampling rate, sampling
at several different rates gives more restrictions over the
originating band $[a,b]$. When sampling at $P$ rates,
$F^1,\ldots,F^p$, the originating band must reside within
$\cap_{i=1}^{P} \cup_n [a_n^i,b_n^i]$.
\section{Reconstruction Method}
In this section we describe an algorithm to reconstruct signals from
an MRS scheme. First, we describe an algorithm for reconstructing
ideal multi-band signals, as defined above. Then we present
modifications \blue{to enable a reconstruction of} signals that may be
contaminated by noise outside of strictly defined bands. While such
signals are not exactly multi-band, we still consider them
multi-band signals provided that \blue{the} noise amplitude is considerably
\blue{lower} than the signal amplitude.

The reconstruction is performed sequentially. In the first step
sets of intervals in the band $[0,F_{\textrm{nyq}}/2]$ that could be the
support of $X(f)$ are identified. These are sets that, when
down-converted at each sampling rate $F^i$, give energy in intervals
in the baseband where significant energy is observed. For each
hypothetical support, the algorithm determines the subsets of the
support that are unaliased in each channel. According to
Eq.~\ref{baseband sum}, for the correct support, the amplitude of
each sampled signal spectrum is proportional to the original signal
spectrum over the unaliased subset of the support. As a result, for
each pair of channels, the amplitudes of the two sampled signal
spectra are proportional to one another over the subsets of the
hypothetical support which are unaliased in both channels. Thus, we
define an objective function that quantifies the consistency between
the different channels over mutually unaliased subsets of the
support. The algorithm chooses the hypothetical support that
maximizes the objective function. The amplitude is reconstructed
from the sampled data on the unaliased subsets of the chosen
hypothetical support. In the last step, the phase of the spectrum of
the originating signal is determined from the unaliased subset of the
chosen hypothetical support.


\subsection{Noiseless signals}

In this subsection we assume that all signals are ideal multi-band
signals. Although what follows applies to more general signals, we
assume that all signals have piece-wise continuous spectrum.

\subsubsection{Reconstruction of the spectrum amplitude}

For each sampled signal $X^i(f)$, we consider the indicator function
\blue{$\mathcal{I}^i(f)$} that indicates over which frequency intervals the
energy of the sampled signal $X^i(f)$ resides. \blue{To} ignore
isolated points discontinuity we define the indicator functions
\blue{$\mathcal{I}^i(f)$} as follows:\blue{
\begin{equation}
\mathcal{I}^i(f)=\left\{
\begin{array}{lll}
1&&\mathrm{for\;all\;} f\in[0,F_{\textrm{nyq}}/2] \text{\;such that for
all\;} \varepsilon>0,
\int_{f-\varepsilon}^{f+\varepsilon} |X^i(f')|^2 df' >0\\
0&&\mathrm{otherwise}.\nonumber
\end{array}
\right.
\end{equation}
}
For piece-wise continuous function, it is simple to show that
\blue{$\mathcal{I}^i(f)=1$} on closed intervals. 

We define the function  \blue{$\mathcal{I}(f)$} as follows:\blue{
\begin{equation}
\mathcal{I}(f)=\prod_{i=1}^{P}\mathcal{I}^i(f),
~~~~~f\in[0,F_{\textrm{nyq}}/2].
\end{equation}
}
Thus, the function \blue{$\mathcal{I}(f)$ equals $1$} over the intersection of all
the up-converted bands of the $P$ sampled signals. We denote the
intervals over which \blue{$\mathcal{I}(f)=1$} by $U_1\ldots U_K$. The
Appendix gives sufficient conditions under which each originating
band \blue{coincides with} one of the intervals $U_1\ldots U_K$. Thus, it remains
to determine which of the $K$ intervals coincide with the
originating intervals.

For each $k=1,2,\cdots,K$ we consider the indicator function
\begin{equation}\label{indicator_Uk}
\mathcal{I}_k(f)=\left\{
\begin{array}{lll}
1&&\mathrm{if\;}f\in U_k\\
0&&\mathrm{otherwise}.
\end{array}
\right.
\end{equation}
It follows immediately from Eq.~\ref{indicator_Uk} that 
\begin{equation}\label{indicator_full}
\mathcal{I}(f)=\sum_{k=1}^K\mathcal{I}_k(f).
\end{equation}
To find which sets of $U_k$ (or \blue{$\mathcal{I}_k(f)$)} match
the originating bands each indicator function \blue{$\mathcal{I}_k(f)$} is
down-converted to the baseband via the formula 
\blue{
\begin{equation}\label{down_conv}
\mathcal{I}^i_k(f)=\mathcal{I}_{[0,F^i/2]}(f)H\left(\sum_{n=-\infty}^{n=\infty}\mathcal{I}_k(f+nF^i)+\mathcal{I}_k(-f+nF^i)\right).
\end{equation}
In Eq.~\ref{down_conv} $\mathcal{I}_{[0,F^i/2]}(f)$ is the indicator function of the closed interval
 $[0,F^i/2]$:
\begin{equation}\label{indicator_interval}
\mathcal{I}_{[0,F^i/2]}(f)= \left\{
\begin{array}{lll}
1&\mathrm{if}&f\in[0,F^i/2]\\
0&\mathrm{otherwise}.&
\end{array}
\right.
\end{equation}
$H(f)$ is the Heaviside step function
\begin{equation}\label{heaviside}
H(f)=\left\{
\begin{array}{lll}
0&\mathrm{if}&f\le 0\\
1&\mathrm{if}&f>0.
\end{array}
\right.
\end{equation}
The Heaviside step function in Eq.~\ref{down_conv} is used to assure
that $\mathcal{I}^i_k(f)$ is an indicator function. In the case in
which the down-conversions of an interval $U_k$ are aliased at some frequency $f$ within
the baseband the argument of the step function is an integer greater than 1.  However,
$\mathcal{I}^i_k(f)=1$.  If, for a frequency $f$ in the baseband there is no signal in any of its replicas; i.e., $F(nF^i\pm f)=0$ for all $n$, then $H(f)=0$.   As a consequence, $\mathcal{I}^i_k(f)=0$ also.  Therefore, the function $\mathcal{I}^i_k(f)$ is equal to one over the
down-conversion of the interval $U_k$ corresponding sampling rate
$F^i$.
}

We consider the power set of $U$,  $\mathcal{P}\{U\}$; i.e., the set
of all subsets of $\{U_1,\cdots,U_K\}$. We denote an element of
$\mathcal{P}\{U\}$ by $\mathcal{U}=\{U_{k_1},\cdots,U_{k_Q}\}$
($0\le Q\le K$). A subset $\mathcal{U}\in\mathcal{P}\{U\}$ is deemed
to be a \textit{support consistent combination} if, for each
sampling rate $F^i$, the down conversion of its intervals matches
the down-converted bands of the corresponding sampled signal. In
terms of indicator functions, we define for each $\mathcal{U}\in
\mathcal{P}\{U\}$ the indicator functions \blue{
\begin{equation}
\mathcal{I}^i_{\mathcal{U}}(f)=\sum_{U_k\in
\mathcal{U}}\mathcal{I}^i_k(f) ~~~~~~~~~f\in[0,F^i/2].
\end{equation}
}
The function \blue{$\mathcal{I}^i_{\mathcal{U}}(f)$} is an indicator
function for the down-conversion of the intervals of $\mathcal{U}$.
Next, we define the objective function
\blue{
\begin{equation}\label{C}
E_1(\mathcal{U})=\sum_{i=1}^P\int_{0}^{F^i/2}|\mathcal{I}^i_{\mathcal{U}}(f)-\mathcal{I}^i(f)|\;\mathrm{d}f.
\end{equation}
}
Support consistent combinations are those $\mathcal{U}$ for which
$E_1(\mathcal{U})=0$.

\begin{figure}[htb]
\hspace{4.5 cm} \includegraphics[angle=-90, width=18cm]{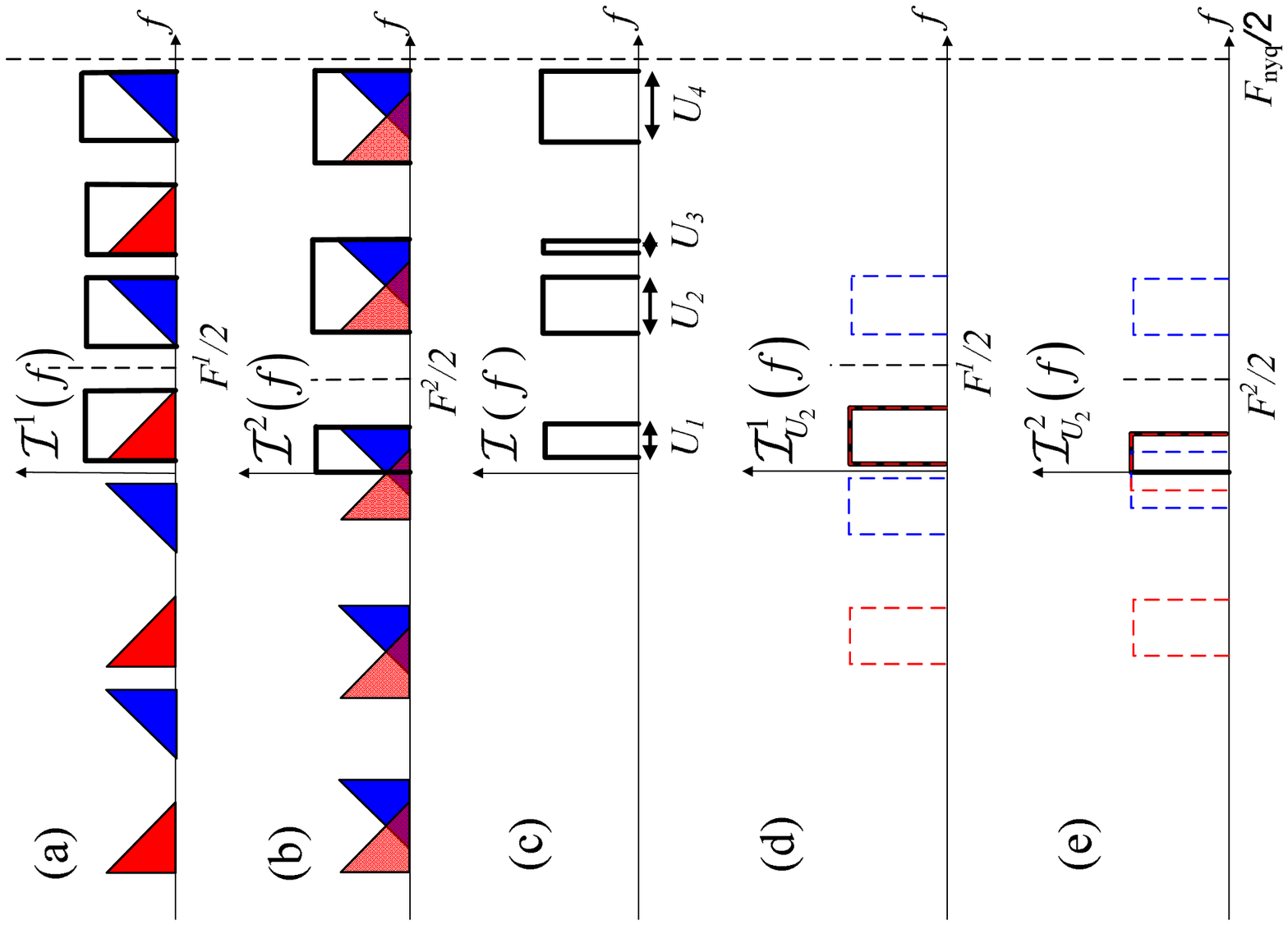}
\caption{\label{illustration2}Illustration demonstrating how support-consistency is
checked. The input of the algorithm is the sampled signals whose
spectra $X^1(f)$ and  $X^2(f)$ are shown Figs.~1 (b) and (c),
respectively; their respective indicator functions
\blue{$\mathcal{I}^1(f)$} and \blue{$\mathcal{I}^2(f)$} are shown in Fig.~2 (a)
and (b). Figure 2 (c) shows the indicator function
\blue{$\mathcal{I}(f)=\mathcal{I}^1(f) \mathcal{I}^2(f)$}. In Figs.~2 (d)
and (e), we check whether the subset $\mathcal{U}=\{U_2\}\in
\mathcal{P}\{U\}$ is support consistent. Figures \ref{illustration2}
(d)and  (e) show the indicator functions for the down-conversion of
$U_2$ at rates $F^1$ and $F^2$: \blue{$\mathcal{I}^1_{U_2}(f)$} and
\blue{$\mathcal{I}^2_{U_2}(f)$}, respectively. The dashed lines illustrate
$U_2$, $-U_2$ and their down-conversions. It is evident that the
functions \blue{$\mathcal{I}^1(f)$} and \blue{$\mathcal{I}^1_{U_2}(f)$} are not
equal. Hence, $\mathcal{U}=\{U_2\}$ is not a support-consistent
combination.}
\end{figure}

Figure \ref{illustration2} illustrates our method for the signal
shown in Fig.~\ref{illustration}. The support of the signal at
positive frequencies, shown in Fig.~\ref{illustration}, consists of
a single interval. Figures \ref{illustration2}(a) and
\ref{illustration2}(b) are graphs of \blue{$\mathcal{I}^1(f)$} and
\blue{$\mathcal{I}^2(f)$}. Figure \ref{illustration2}(c) is a graph of
\blue{$\mathcal{I}(f)$}. The function \blue{$\mathcal{I}(f)$} is equal to one over
four intervals $U_1, \ldots, U_4$. Each combination of these four
intervals must be checked for support consistency. In the example
illustrated in Fig.~\ref{illustration2}, we check whether the subset
$\mathcal{U}=\{U_2\}\in \mathcal{P}\{U\}$ is support consistent.
Figures \ref{illustration2} (d) and (e) show the indicator functions for
the down-conversion of $U_2$ at rates $F^1$ and $F^2$:
\blue{$\mathcal{I}^1_{U_2}(f)$} and \blue{$\mathcal{I}^2_{U_2}(f)$}, respectively.
The dashed lines illustrate $U_2$, $-U_2$, and their
down-conversions. It is evident that the functions
\blue{$\mathcal{I}^1(f)$} and \blue{$\mathcal{I}^1_{U_2}(f)$} are not equal.
Hence, $\mathcal{U}=\{U_2\}$ is not a support-consistent
combination.

Amongst all support consistent combinations $\mathcal{U}$, it is
necessary to identify the one that exactly matches the originating
bands. For this purpose, we introduce two additional objective
functions. The support consistent combination $\mathcal{U}$ that
optimizes these function is deemed to be the correct one.

Amongst support-consistent combinations, amplitude consistent
combinations are defined by the amplitudes of the sampled
signals at unaliased intervals. Let
$\mathcal{U}=\{U_{j_1},\cdots,U_{j_m}\}$ be a support consistent
combination. Denote the union of  all intervals in
$\bigcup_{n=1}^mU_{j_n}$ that are unaliased when down-converted at
rate $F^{i}$ by $\Sigma^{i}_{\mathcal{U}} \subset
\bigcup_{n=1}^mU_{j_n}$. For the correct choice of $\mathcal{U}$, at
a frequency $f$ that is unaliased when sampled at rates $F^{i_1}$
and $F^{i_2}$ ( $f\in
\Sigma_{\mathcal{U}}^{i_1}\cap\Sigma_{\mathcal{U}}^{i_2}$), the
functions $|X^{i_1}(f)|/F^{i_1}$ and $|X^{i_2}(f)|/F^{i_2}$ must be
equal. Accordingly, we define a second objective function:
\begin{equation}\label{E1}
E_2(\mathcal{U})=\sum_{i_1\ne
i_2}\int_{\Sigma_{\mathcal{U}}^{i_1}\cap\Sigma_{\mathcal{U}}^{i_2}}(|X^{i_1}(f)|/F^{i_1}-|X^{i_2}(f)|/F^{i_2})^2\;\mathrm{d}f.
\end{equation}
For the correct $\mathcal{U}$, the objective function
$E_2(\mathcal{U})$ must equal zero. A support-consistent combination
$\mathcal{U}$ for which $E_2(\mathcal{U})=0$ is said to be amplitude
consistent.

Unfortunately, there may be more than one amplitude-consistent
combination. This is the case, for example, when for all $i_1$ and
$i_2$, $\Sigma^{i_1}_{\mathcal{U}}\cap \Sigma^{i_2}_{\mathcal{U}}$
is empty. In such cases, the objective function $E_2(\mathcal{U})$
cannot be sufficient to identify the correct $\mathcal{U}$. Thus, we
introduce a third objective function $E_3(\mathcal{U})$. This
function favors options in which the integrals in Eq.~\ref{E1} are
calculated over large sets. The third objective function is defined
by
\begin{equation}\label{E2}
E_3(\mathcal{U})=\sum_{i_1\ne i_2}
\lambda(\Sigma^{i_1}_{\mathcal{U}}\cap\Sigma_{\mathcal{U}}^{i_2}),
\end{equation}
where
$\lambda(\Sigma^{i_1}_{\mathcal{U}}\cap\Sigma_{\mathcal{U}}^{i_2})$
is the Lebesgue measure of
$\Sigma^{i_1}_{\mathcal{U}}\cap\Sigma_{\mathcal{U}}^{i_2}$. The
amplitude-consistent combination that maximizes $E_3(\mathcal{U})$
is deemed to be the correct one. In the rare case that
$E_3(\mathcal{U})$ is maximized by more than one
amplitude-consistent combination, the outcome of the algorithm is
not determined.

After the optimal $\mathcal{U}=\{U_{j_1},\cdots,U_{j_m}\}$ is chosen, the
amplitude of the signal is reconstructed from the samples. We define
the function $r(f)$ as the number of sampled signals which are
unaliased at the frequency $f$: \blue{$r(f)=\sum_{i=1}^{P}
\mathcal{I}_{\Sigma^{i}_{\mathcal{U}}}(f)$}, where
\blue{$\mathcal{I}_{\Sigma^{i}_{\mathcal{U}}}(f)$} is the indicator
function of the interval ${\Sigma^{i}_{\mathcal{U}}}$, defined
similarly to Eq.~\ref{indicator_interval}. For each $f$ within the 
detected originating bands, i.e. $f\in\bigcup_{n=1}^m U_{j_n}$, if
$r(f)>0$, we reconstruct the corresponding amplitude of the spectrum
at $f$ from the sampled signals by
\begin{equation} \label{full}
X_{\mathcal{U}}(f)=\frac{1}{r(f)}\sum_{i=1}^{P}
\frac{|X^{i}(f)|\mathcal{I}_{\Sigma^{i}_{\mathcal{U}}}(f)}{F^{i_n}}. 
\end{equation}
In words, for each frequency $f$ that is unaliased in at least one
channel, the signal amplitude is averaged over all the channels that
are not aliased at $f$. For all other frequencies, notably those
that alias in all sampling channels, $X_{\mathcal{U}}(f)$ is set to
equal zero.

\subsubsection{Reconstruction of the spectrum phase}

\blue{The spectrum of a signal can be expressed as $X(f)=|X(f)|\exp\{j \arg[X(f)]\}$. In the previous section we described how to reconstruct the amplitude $|X(f)|$ from the signal's sampled data. In this section we describe a method of reconstructing the phase $\arg[X(f)]$.}
If the time offsets $\Delta^{i}$ \blue{of Eq.~\ref{baseband sum}}  \blue{were} known a priori, reconstructing
the phase \blue{would be} trivial. The reconstruction  \blue{in this case could} be performed by using a
variant of Eq.~\ref{full} \blue{with} $|X^{i_n}(f)|$  replaced by
$X^{i_n}(f)\exp(-2\pi f \Delta^{i_n})$. This would yield a full
reconstruction of the signal (phase and amplitude). However, because
of the lack of synchronization between the channels, the time
offsets \blue{$\Delta^i$} are not known a priori.  \blue{Consequently}, it is more
difficult to reconstruct the phase. After identifying the signal bands, we can calculate the difference\blue{s $\Delta^{i_1}-\Delta^{i_2}$} between two different time offsets.
This is sufficient to enable the reconstruction of the phase of the signal spectrum
up to a single linear phase factor.

The difference between two time offsets $\Delta^{i_1}$ and
$\Delta^{i_2}$ can be calculated directly in the case that
$\Sigma_{\mathcal{U}}^{i_1}\cap \Sigma_{\mathcal{U}}^{i_2}$ contains
at least one finite interval. In this interval the phase of
$X^{i_1}(f)/X^{i_2}(f)$ satisfies the following equation:
\begin{equation}\label{phase_diff}
\arg[X^{i_1}(f)/X^{i_2}(f)]=2\pi f (\Delta^{i_1}-\Delta^{i_2}) +2\pi
k,~~~~~~\text{for some integer}~k
\end{equation}
The left side of Eq.~\ref{phase_diff} is determined by the sampled
data. By performing a linear fit we calculate the difference
between the two offsets $\Delta^{i_1}$ and $\Delta^{i_2}$. We do
this for all pairs of offsets for which
$\Sigma_{\mathcal{U}}^{i_1}\cap \Sigma_{\mathcal{U}}^{i_2}$ contains
at least one finite interval.

There may exist cases in which there exist $i_1$ and $i_2$ such that
$\Sigma_{\mathcal{U}}^{i_1}\cap
\Sigma_{\mathcal{U}}^{i_2}$ does not contain one finite interval but for
which $\Delta^{i_1}-\Delta^{i_2}$
can still be calculated. For example, in the case of three offsets
$\Delta^{i_1}$, $\Delta^{i_2}$ and $\Delta^{i_3}$, if one can
calculate $(\Delta^{i_1}-\Delta^{i_2})$ and
$(\Delta^{i_2}-\Delta^{i_3})$, then $(\Delta^{i_1}-\Delta^{i_3})$
can also be calculated by  simple algebra. If there exist ${i_{n}}, \ldots
{i_{m}}$, such that for each ${n}\leq k\leq {m-1}$,  $\Sigma_{\mathcal{U}}^{i_k}\cap
\Sigma_{\mathcal{U}}^{i_{k+1}}$ contains at least one finite  
interval, then we say that $i_{n}$ and ${i_{m}}$ are \emph{phase
connected} and denote this by ${i_n}\sim {i_m}$.
If  $i\sim j$, then difference between the two offsets $\Delta^{j}-\Delta^{i}$ can be
calculated.  In the
case $\Sigma_{\mathcal{U}}^{i}$ does not contain any finite
intervals, we define $\Delta^i\sim \Delta^i$. It is clear that
$\sim$ is an equivalence relation \cite{equivalence} and thus
partitions the $\Delta^i$ into equivalence classes.

For each $\Delta^{i_1}$ and $\Delta^{i_2}$ in the same class, one
can calculate their difference. One can obtain a full
reconstruction of the phase if there exists  one class
$\mathcal{C}$ such that each originating frequency is unaliased in
at least one channel belonging to $\mathcal{C}$; i.e, there exist a
class $\mathcal{C}=\Delta^{i_n}\ldots \Delta^{i_m}$, such that
$\bigcup_{k=n}^{m} \Sigma_{\mathcal{U}}^{i_k}=\bigcup_{k=1}^Q
U_{j_k}$, where $\mathcal{U}=\{U_{j_1},\cdots,U_{j_Q}\}$.

\subsection{Physical signals}
\blue{To} sample realistic signals (i.e., not strictly multi-band
and in the presence of noise), the algorithm needs to be adjusted.
In this subsection we describe adjustments to our algorithm to
overcome the noise. The algorithm requires five new parameters. In
section 5, we give examples of reconstructing signals contaminated
by strong noise. In those examples, the success of the
reconstruction does not depend on the exact choice of the five
parameters.

In the presence of noise, the definition of the support of the
sampled signals must be adjusted. First, a small $\xi$ is chosen.
Then, a small positive threshold value $T$ is chosen.   The
indicator function \blue{$\mathcal{I}^i(f)$} is then redefined as follows: \blue{
\begin{equation}\label{indicator}
\mathcal{I}^i(f)=\left\{
\begin{array}{lll}
1&&\mathrm{if\;}f\in[0,F_{\textrm{nyq}}/2]\;\mathrm{and}\;\frac{1}{2\xi}\int_{f-\xi}^{f+\xi}|X_\alpha(f')|df'>T\\
0&&\mathrm{otherwise}.
\end{array}
\right .
\end{equation}
}
The choice of the threshold $T$ depends on the average noise level.

When reconstructing physical signals, it is not reasonable to expect
$E_1(\mathcal{U})$ to equal 0 for any combination $\mathcal{U}$. An
initial adjustment is to require that $E_1(\mathcal{U})<b$ for some
positive $b$. The shortcoming of this condition is that the
threshold $b$ does not depend on the signal. \blue{To} make the
threshold to depend on the signal in a simple way, we introduce the
following condition:
\begin{equation} \label{support_con}
E_1(\mathcal{U})<a \min_\mathcal{U}\left[E_1(\mathcal{U})\right]+b
\end{equation}
where $a \ge 1$ is a chosen parameter. The parameters $a$ and $b$
control the tradeoff between the chance of success and runtime. If
$a$ and $b$ are too small, the correct subset $\mathcal{U}$ may not
be included in the set of support constituent combinations.  On the
other hand, if $a$ and $b$ are too large, then the number of
support-consistent combinations may be large. This results in a slow
run time.

Finally, we make two modifications to the objective function
$E_3(\mathcal{U})$. We replace the length of the mutually unaliased
intervals by a weighted energy of the sampled signals in these
interval. The objective function $E_3(\mathcal{U})$ is replaced with
$\widehat{E}_3$:
\begin{equation}\label{E2_new}
\widehat{E}_3(\mathcal{U})=\sum_{i_1\ne
i_2}\int_0^{F_{\textrm{nyq}}/2}\left|\frac{X^{i_1}(f)}{F^{i_1}}\right|^2{W}_{i_1,i_2}(f,\mathcal{U})\;\mathrm{d}f,
\end{equation}
where ${W}_{i_1,i_2}(f,\mathcal{U})$ is a weight function. The
weight function favors combinations in which the sampled signals are
similar in mutually unaliased internals and is defined in the
following.

We first note that for each two channels $i_1$ and $i_2$, the
intersection of their non-aliased supports
($\Sigma^{i_1}_{\mathcal{U}}\cap\Sigma_{\mathcal{U}}^{i_2}$) is a
union of a finite number of disjoint intervals $V_1^{i_1,i_2},\cdots
V_R^{i_1,i_2}$. We define
\begin{equation}
\mu^k_{i_1,i_2}(\mathcal{U})=\frac{\int_{V_k^{i_1,i_2}}|\;|X^{i_1}(f)|/F^{i_1}-|X^{i_2}(f)|/F^{i_2}\;|\;\mathrm{d}f}{\int_{V_k^{i_1,i_2}}|\;|X^{i_1}(f)|+|X^{i_2}(f)|\;|\;\mathrm{d}f}.
\end{equation}
Finally, we define the weight function:
\blue{
\begin{equation} \label{W_smooth}
W_{i_1,i_2}(f)=\sum_k\exp[-\rho{\mu}^k_{i_1,i_2}(\mathcal{U})]{\mathcal{I}}_{V_k^{i_1,i_2}}(f),
\end{equation}
}
where $\rho$ is a chosen positive constant and
\blue{$\mathcal{I}_{V_k^{i_1,i_2}}(f)$} is the indicator function of the
interval $V_k^{i_1,i_2}$. The parameter $\rho$ is chosen according
to an assumed signal to noise ratio (SNR). When the SNR is lower, in
order to accept higher errors $\rho$ is chosen to be smaller. In the
case of a noiseless signal and an amplitude-consistent
${\mathcal{U}}$, each $\mu^k_{i_1,i_2}$ vanishes. Therefore, in this
case, each element in the sum on the right-hand side of
Eq.~\ref{E2_new} gives the energy of the signal over
$\Sigma^{i_1}_{\mathcal{U}}\cap\Sigma_{\mathcal{U}}^{i_2}$. In all
other cases, the energy in each interval  $V_k^{i_1,i_2}$ is
weighted according to the relative error between $X^{i_1}(f)$ and 
$X^{i_2}(f)$ over $V_k^{i_1,i_2}$.

Since in the case of noisy signals, neither $E_1(\mathcal{U})$ nor
$E_2(\mathcal{U})$ vanishes for the combination which corresponds to
the originating bands, both  $E_1(\mathcal{U})$ and
$E_2(\mathcal{U})$ should be considered in the final step of
choosing the best combinations. Accordingly, we define the following
objective function $E_\text{tot}(\mathcal{U})$:
\begin{equation}
E_\text{tot}(\mathcal{U})=-\frac{E_1(\mathcal{U})}{\min_{\mathcal{U}}\left\{E_1(\mathcal{U})\right\}}
-\frac{E_2(\mathcal{U})}{\min_{\mathcal{U}}\left\{E_2(\mathcal{U})\right\}}
+\frac{\widehat{E}_3(\mathcal{U})}{\min_{\mathcal{U}}\left\{\widehat{E}_3
(\mathcal{U})\right\}}
\end{equation}
for all $\mathcal{U}$ such that
${\min_{\mathcal{U}}\left\{E_1(\mathcal{U})\right\}},
{\min_{\mathcal{U}}\left\{E_2(\mathcal{U})\right\}},
{\min_{\mathcal{U}}\left\{E_3(\mathcal{U})\right\}} \neq 0$. Amongst
all such combinations that also satisfy Eq.~\ref{support_con}, the
one that gives the maximum value of $E_\text{tot}(\mathcal{U})$ is
deemed to be correct. In cases in which either
$\min_{\mathcal{U}}\left\{E_1(\mathcal{U})\right\}$,
$\min_{\mathcal{U}}\left\{E_2(\mathcal{U})\right\}$ or
$\min_{\mathcal{U}}\left\{E_3(\mathcal{U})\right\}$ equals zero for
a certain combination $\mathcal{U}$, the maximum of
$\widehat{E}_3(\mathcal{U})$ is chosen as the solution.

%
%
%

To reconstruct the phase, the only change made is in how the
difference between the offsets is calculated. Equation
\ref{phase_diff} holds for all the disjoint intervals
$V_k^{i_1,i_2}\in \Sigma_{\mathcal{U}}^{i_1}\cap
\Sigma_{\mathcal{U}}^{i_2}$. Accordingly, we perform the linear fit
for each intervals, and obtain a certain value for
$\Delta^{i_1}-\Delta^{i_2}$. Each value is weighted by the length of
its respective $V_k^{i_1,i_2}$. These weighted values are averaged.
The result is an estimate for $\Delta^{i_1}-\Delta^{i_2}$. This
averaging procedure may increase the accuracy in the estimate of
$\Delta^{i_1}-\Delta^{i_2}$.

\section{Complexity considerations}
In this section we discuss considerations used to reduce the computational complexity of our
algorithm. Choosing a subset $\mathcal{U}\in \mathcal{P}\{U\}$
involves calculating three objective functions. We explain why
eliminating possibilities through the use of $E_1(\mathcal U)$
alone can significantly reduce runtime.

In the first step of the algorithm, we find support consistent
combinations by calculating the objective function
$E_1(\mathcal{U})$ for elements in $\mathcal{P}\{U\}$. Assuming the
largest element in $\mathcal{P}\{U\}$ contains $K$ intervals, and
that the signal is composed of up to $N$ bands in $[0,F_{\textrm{nyq}}/2]$,
the number of elements in $\mathcal{P}\{U\}$ that one needs to check
is equal to
\begin{equation} \label{complexity}
\sum_{n=1}^{N} \left( \begin{array}{cc}
         K \\ n
       \end{array}
 \right) .
 \end{equation}
In the case $N \approx K$,  the complexity is approximately
$O(2^N)$. When $N/K\ll 1 $, the last term in Eq.~\ref{complexity}
number of options to be checked is approximately equal to
$O(K^N/N!)$.

The complexity of checking a single option out of $\mathcal{P}\{U\}$
for support consistency (Eq.~\ref{C}) is $O(1)$ and it does not depend
on the number of points used to discretize the spectrum. By
contrast, the complexity of checking such an option for amplitude
consistency (Eqs.~\ref{E1} and \ref{E2}) is of the order of the
number of points used to represent the spectrum. This is a major
reason for using the support-consistency criterion to narrow down
the number of options needed to be checked for amplitude
consistency. The amplitude
consistency is calculated only for
support-consistent options, which are in general much fewer than
what is prescribed by Eq.~\ref{complexity}.

\section{Numerical Results}
This section describes results of our numerical simulations. The
simulations were carried out in the two cases considered in the
previous sections: i) ideal multi-band signals and ii) noisy
signals. In all our examples, the number of channels $P$ was set equal
to three, $P=3$.

In all our simulations, the number the bands in $[0,F_{\textrm{nyq}}/2]$
equals $N$, where $N\leq 4$. Unless stated otherwise the band number
refers to the number of bands in the non negative frequency region
$[0,F_{\textrm{nyq}}/2]$. Using the notations in Eq.~\ref{multi_band}, each
signal in each band is given by
\begin{equation}\label{signal_cos}
S_n(f)=\left\{
\begin{array}{lll}
A_n\cos[\pi(f-f_n)/B_n]&&\mathrm{if\;} 2|f-f_m|/B_n<1\\
0&&\mathrm{otherwise},
\end{array}
\right.
\end{equation}
where $B_n$ is the spectral width of the $n$th band, $f_n$ is its
central frequency, and $A_n$ is the maximum amplitude. The total
spectral measure of the signal support equals
$\Sigma_x=2\sum_{n=1}^{N} B_n$, and the minimal sampling rate is
equal to $2\Sigma_x$ \cite{Moshiko}, twice the Landau rate. In each
simulation, all the bands had the same width, i.e.
$B_n=\Sigma_x/(2N)$. The amplitudes $A_n$ were chosen independently
from a uniform distribution on $[1,1.2]$. The central frequencies
$f_n$ were also chosen independently from a uniform distribution on
the region $[0,F_{\textrm{nyq}}/2]$. We eliminated cases in which there was
an overlap between any two different bands. The time offsets,
$\Delta^i$ were chosen independently from a uniform distribution on
$[0,1/B_n]$.

In each of the simulations, we set $B=800$ MHz and $40\leq
F_{\textrm{nyq}}\leq 76$ GHz. This choice of parameters is consistent with 
previous optical sampling experiments \cite{Avi}. The sampling rates
were chosen as $F^1=3.8 F_0$, $F^2=4 F_0$, and $F^3=4.2 F_0$, where
the value $F_0$ varied between simulations. These sampling rates
were chosen such that, for each pair of sampling rates ($F^i$,
$F^j$), the functions \blue{$\mathcal{I}^i(f),~\mathcal{I}^j(f)$} do not
have a common multiple smaller than $F_{\textrm{nyq}}$. This condition is satisfied
for all $F_0>F_{\textrm{nyq}}/76$.    

\blue{To} obtain an exact reconstruction, the resolution in which
the spectrum is represented $\Delta f$ should be such that the
discretization of the originating baseband downconverts exactly to
the discretization grid in each baseband. This condition is
satisfied when $F^i/\Delta f$ ($i=1,2,3$) is an integer. In our
examples, we used a spectral resolution $\Delta f= 0.8$MHz for all
the channels.

The use of the same spectral resolution for all channels is not only
convenient for implementation of our algorithm, but it also
compatible with the operation of the sampling system used in our experiments \cite{Avi}. In the
implementation of the sampling system, an optical system performs
the down conversion of the signal by multiplying it by a train of
short optical pulses. In each channel a different
repetition rate of the optical pulse train is used. The sampled
signal in each channel is then converted into an electronic signal
and passed through a low-pass filter which rejects all frequencies
outside the baseband. The $P$ filtered sampled signals have a
limited bandwidth. These signals are sampled once more, this time at
a constant rate, using $P$ electronic analog to digital converters.
The use of the optical system allows the use of electronic analog to
digital converters whose bandwidth is significantly lower than the
bandwidth of the multi-band signal \cite{Avi}. Because the signals
at the basebands are sampled with the same time resolution and have
the same number of samples, their spectra, which are obtained using
the Fast Fourier Transform, have the same spectral resolution.

In the first set of simulations we increased the signal bandwidth,
without changing the sampling rates. We used two performance criteria:
correct detection of the originating bands and exact reconstruction
of the signal. As to the first criterion, we required only that the
spectral support of the signal be detected without an error. As to
the second criterion, we required that the signal spectrum (phase and amplitude)
be fully and exactly reconstructed without any error. Because the second
criterion concerns exact reconstructions, in the case that the
algorithm failed to reconstruct the signal at even a single
frequency, it was considered to have failed the second criterion.

\begin{figure}[htb]
\begin{center}
$\begin{array}{c@{\hspace{1in}}c} \multicolumn{1}{l}{} &
    \multicolumn{1}{l}{} \\ \hspace{-1.5 cm} \hbox{\includegraphics[angle=-90,width=9cm]{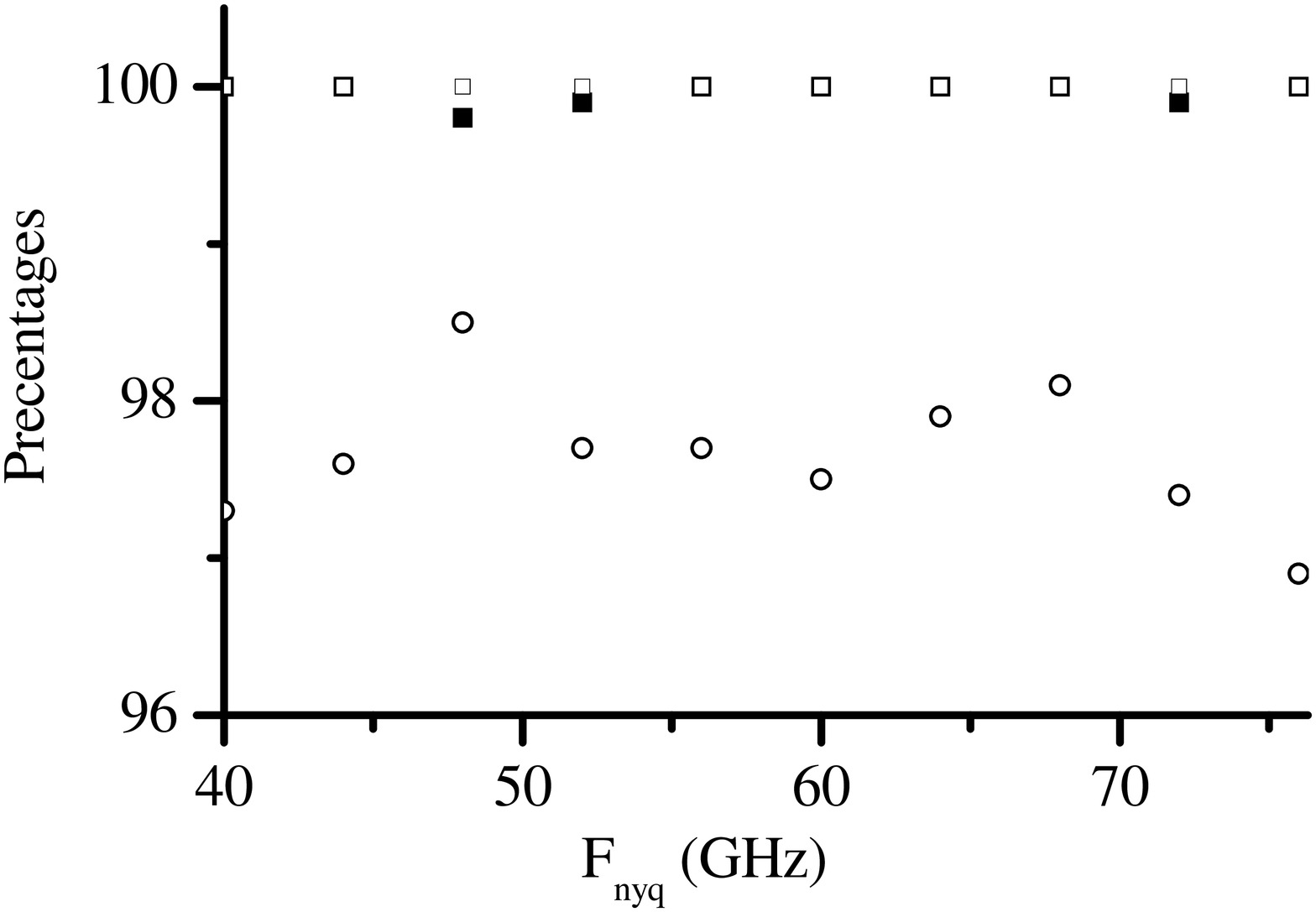}}
    & \hspace{-2 cm}
    \hbox{\includegraphics[angle=-90,width=9cm]{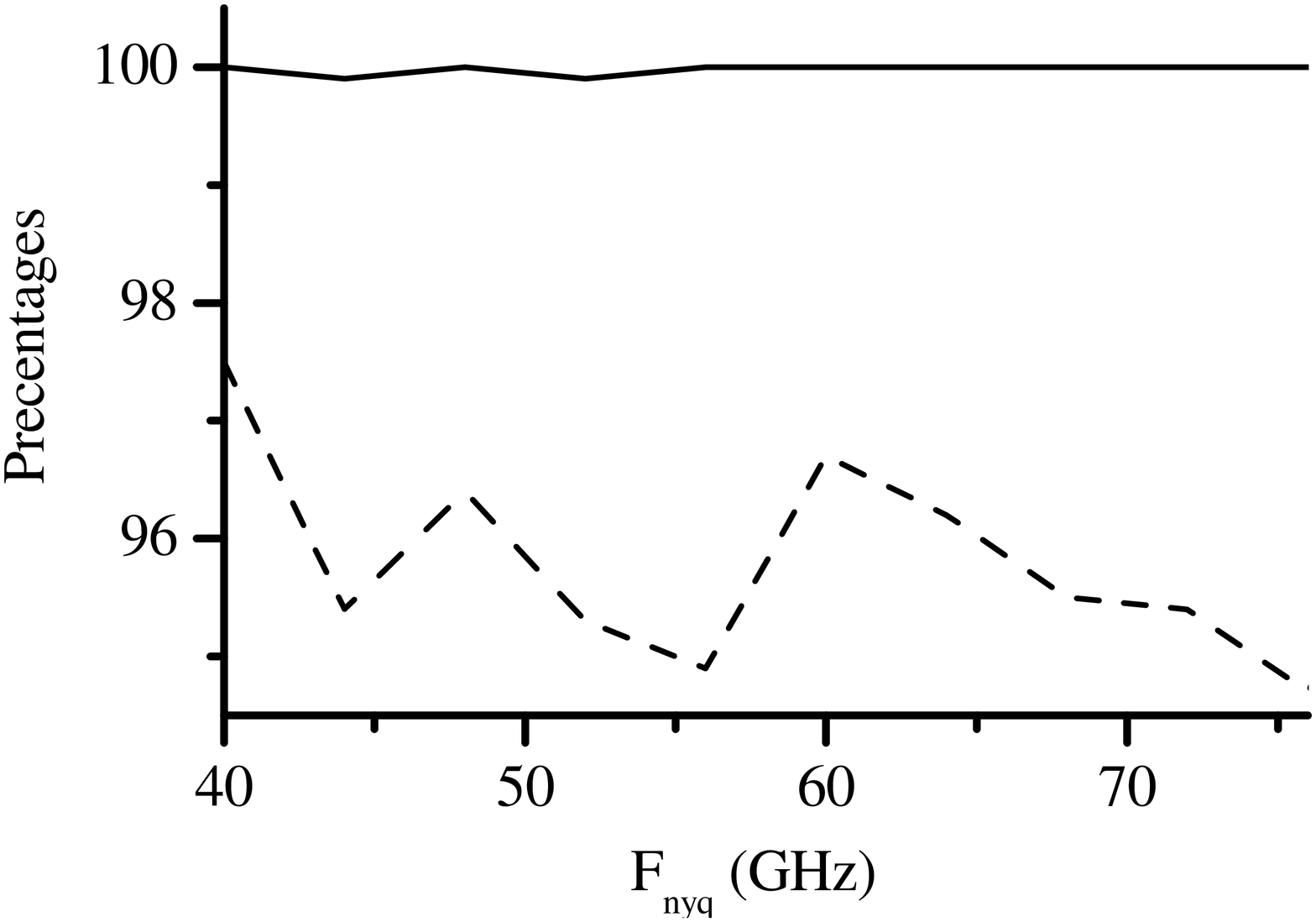}} \\
    \vspace{-1cm}
\mbox{\bf (a)} & \mbox{\bf (b)}
\end{array}$
\end{center}
   \vspace{1cm}  \caption{\label{Nyq}Success percentage for the first set of simulations with
$F_0=1$ GHz as a function of Nyquist rate.
In Fig.~\ref{Nyq} (a), the percentage of a correct band
detection is shown by the squares. The full reconstruction
percentage is shown by circles. The open circles and squares
represent the results obtained when the assumed maximum number of
positive bands equals 3. The dark circles and squares represent the
cases in which the maximum assumed positive band number equals 4.
Figure \ref{Nyq} (b) shows the
band-detection percentage (solid curve) and reconstruction
percentages (dashed curve) in the case that both the maximum number
of originating and assumed positive bands equals 4.}
\end{figure}

We chose $F_0=1$ GHz. This
corresponds to a total sampling rate $F_\text{tot}=F^1+F^2+F^3$ which equals 15 times the Landau rate (7.5 the
minimum possible rate).  The statistics were obtained by averaging over 1000
runs.  Figures \ref{Nyq} (a) and (b) show the results for signals with 3
and 4 positive bands, respectively, as a function of the Nyquist rate. In Fig.~\ref{Nyq} (a), the
percentage of a correct band detection is shown by the squares,
whereas the full reconstruction percentage is shown by circles. The
open circles and squares represent the results obtained when the
maximum number of bands assumed by the algorithm was 3, and
the dark circles and squares represent the cases in which the
maximum assumed band number was equal to 4. Figure
\ref{Nyq} (b) shows the band-detection percentage (solid curve)
and reconstruction percentages (dashed curve) in the case that both
the maximum number of originating and assumed bands is 4. The figures
show that both the success percentages were high and were
not significantly dependent on the Nyquist rate of the signal
or on the number of assumed bands.
\begin{figure}[htb]
\centering\includegraphics[angle=-90,width=12cm]{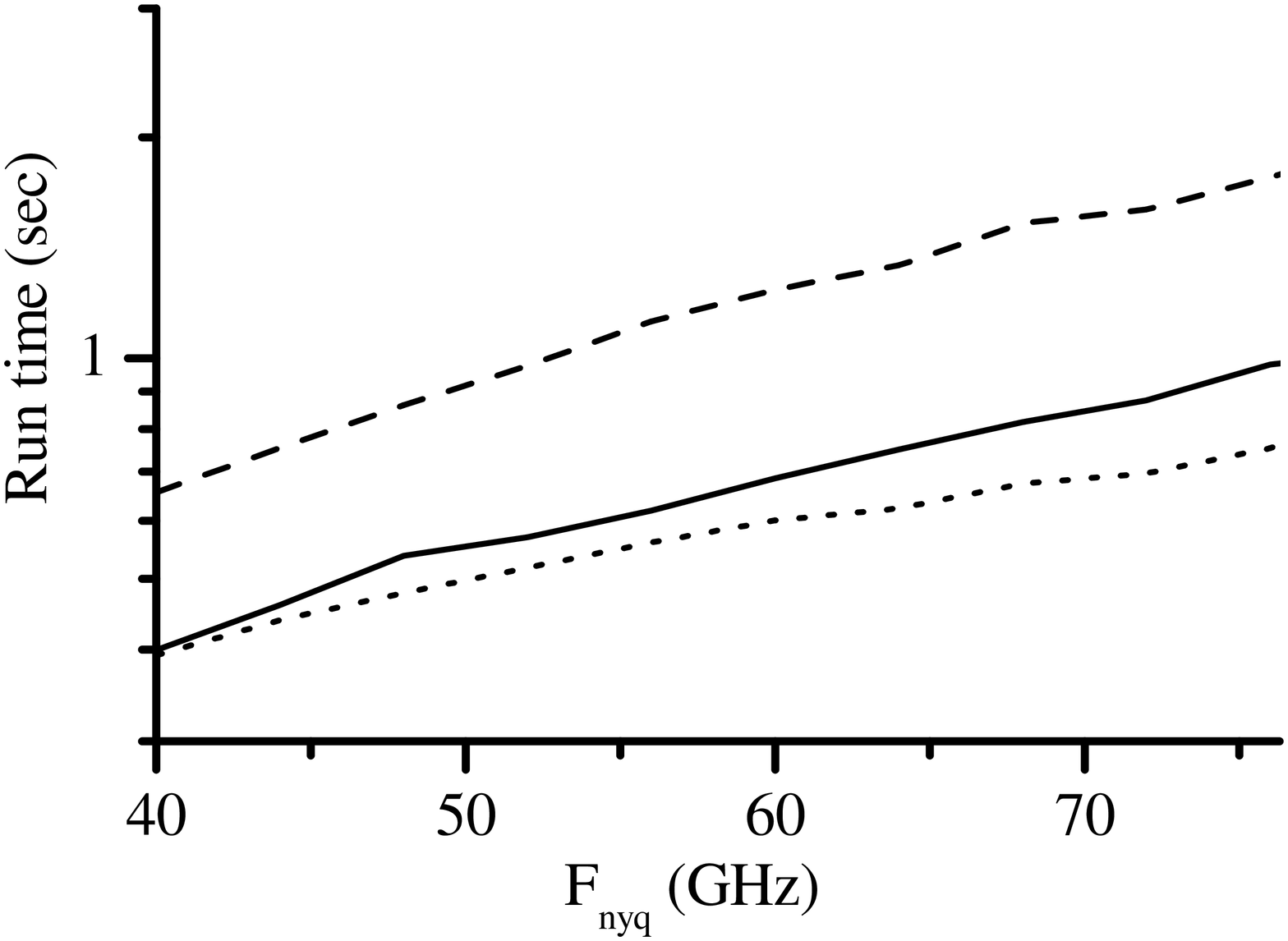}
\caption{\label{time2}The run time for the second set of simulations as a function
of the Nyquist rate. The results in the case of 4 input positive
bands with assumed number of positive bands equals 4 is shown in the
solid curve. The results in the case of 3 input positive bands is
shown with the dotted curve in the case of 3 assumed positive bands
and with the dashed curve in the case of 4 assumed positive bands.}
\end{figure}

Figure \ref{time2} shows the average run time as a function of the Nyquist rate.
The results in the case of 4 input  bands in which the assumed maximum number of
bands is 4 is shown in the solid curve. The results in the case of 3
input bands is shown with the dotted curve in the case of 3
assumed  bands and with the dashed curve in the case of 4
assumed bands. The results show that while an increase in the Nyquist rate
does not significantly affect the reconstruction statistics, it results in an
increase in run time.

\begin{figure}[htb]
\begin{center}
$\begin{array}{c@{\hspace{1in}}c} \multicolumn{1}{l}{} &
    \multicolumn{1}{l}{} \\ \hspace{-1.5 cm} \hbox{\includegraphics[width=9cm]{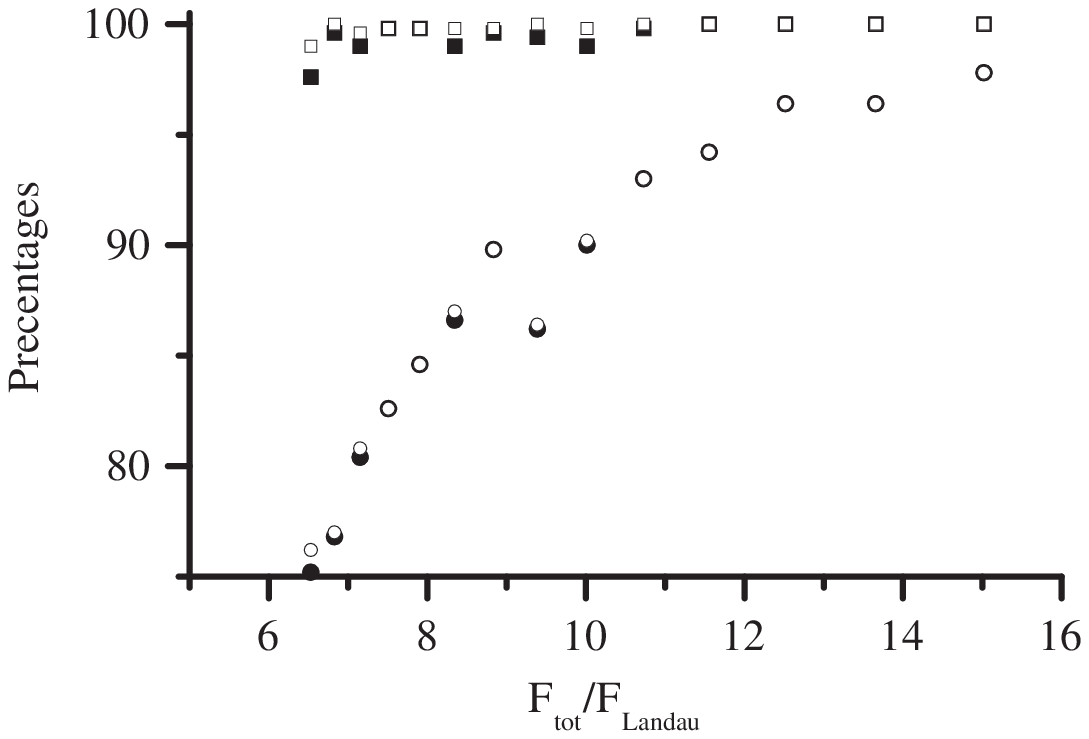}}
    & \hspace{-2 cm}
    \hbox{\includegraphics[width=9cm]{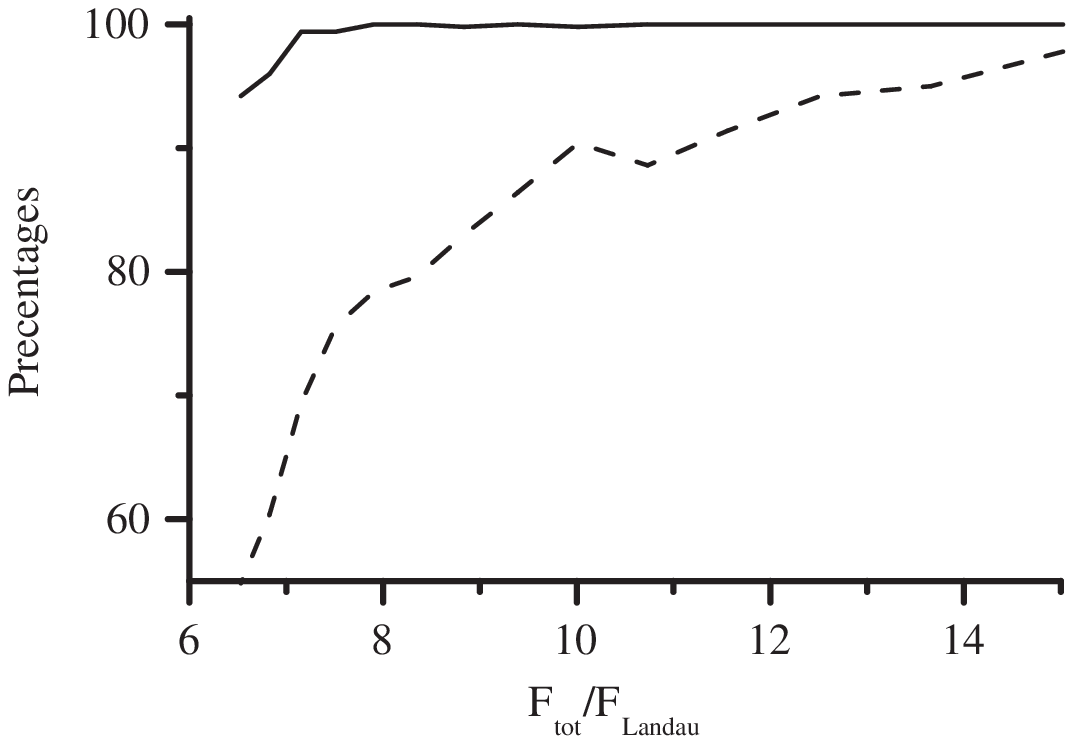}} \\
    \vspace{-1cm}
\mbox{\bf (a)} & \mbox{\bf (b)}
\end{array}$
\end{center}
   \vspace{1cm}  \caption{\label{rate}Success percentage for the first set of simulations as a
function of the sum of the sampling rates divided by the Landau
rate. As in Fig.~\ref{Nyq}, in
Fig.~\ref{rate} (a), the percentage of a correct band detection is
shown by the squares. The full reconstruction percentage is shown by
circles. The open circles and squares represent the results obtained
when the assumed maximum number of positive bands equals 3. The dark
circles and squares represent the cases in which the maximum assumed
positive band number equals 4.
Figure \ref{rate} (b) shows the band-detection percentage (solid
curve) and reconstruction percentages (dashed curve) in the case
that both the maximum number of originating and assumed positive
bands equals 4.}
\end{figure}

\begin{figure}[htb]
\centering\includegraphics[width=12cm]{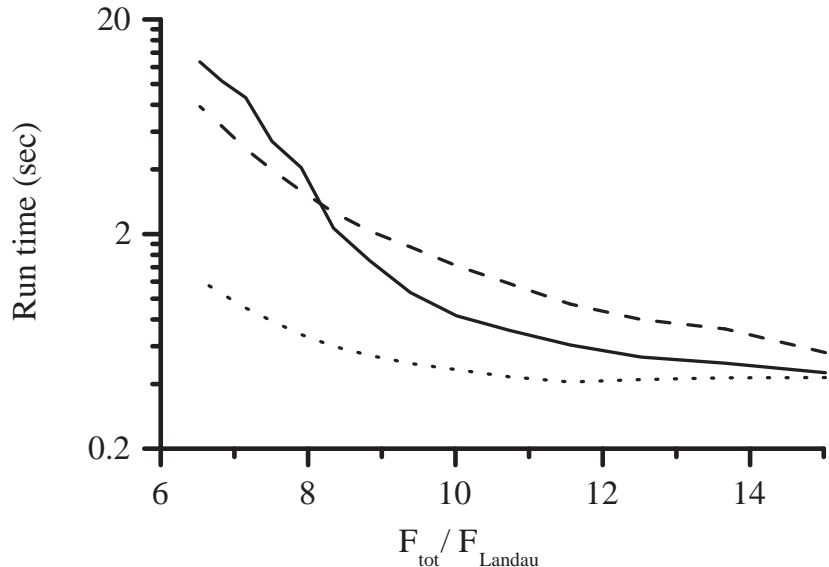}
\caption{\label{time1}The run time for the first set of simulations as a function
of the sum of the sampling rates divided by the Landau rate in the
cases of signals with 4 and 3 positive bands. The results in the
case of 4 input positive bands with assumed number of positive bands
equals 4 is shown in the solid curve. The results in the case of 3
input positive bands is shown with the dotted curve in the case of 3
assumed positive bands and with the dashed curve in the case of 4
assumed positive bands.}
\end{figure}

In the second set of simulations, we measured the performance of our
algorithm as a function of $F_0$. The Nyquist rate used in the
simulation was $F_{\textrm{nyq}}=40$ GHz. For each choice of $F_0$, the
statistics were obtained by averaging over 500 runs. The results did
not change significantly when the averaging was performed over 1000
runs. The simulation was run for the same number of originating
bands and assumed bands as in the first set of simulations. Figures
\ref{rate} (a) and (b) show the success percentages for signals with
3 and 4 bands, respectively, and Fig.~\ref{time1} shows the average
run time. The two success percentages and run time are shown as a
function of the total sampling rate $F_\text{tot}$ divided by the
Landau rate, $F_\text{Landau}=800$ MHz. The symbols used in
Figs.~\ref{rate} (a) and (b) and Fig.~\ref{time1} correspond to
those used in Figs.~\ref{Nyq} (a) and (b) and Fig.~\ref{time2}
respectively.

The results shown in Figs.~\ref{rate} (a) and (b) demonstrate that,
in all the cases that we checked, the average percentage of
successful band detection was over 99.5\% for sampling frequencies
above 8 times the Landau rate. The reconstruction percentages were
lower than these band-detection percentages and were also much more
affected by the sampling rate and by the number of originating
bands. As expected, the run time increases
dramatically with reduction of the sampling rate and also increases
with the assumed maximum number of  bands. We ran similar
simulations with different numbers of originating bands and
different numbers of assumed bands. The trends were similar.

%

In the final set of simulations, the signals are noisy.  We added to
the originating signal white Gaussian noise in the band
$[-F_{\textrm{nyq}}/2,F_{\textrm{nyq}}/2]$, where $F_{\textrm{nyq}}=40$ GHz. We denote by
$\sigma$ the standard deviation of the Gaussian noise in the
pre-sampled signal. Upon sampling the signal at rate $F^i$, the
standard deviation of the
noise increases to $\sigma^i=\sigma \sqrt{\lceil F_{\textrm{nyq}}/F^i \rceil }$ owing to 
aliasing of the noise, where $\lceil x \rceil$ equals the smallest integer greater or equal to $x$.
\begin{figure}[htb]
\centering\includegraphics[angle=-90,width=12cm]{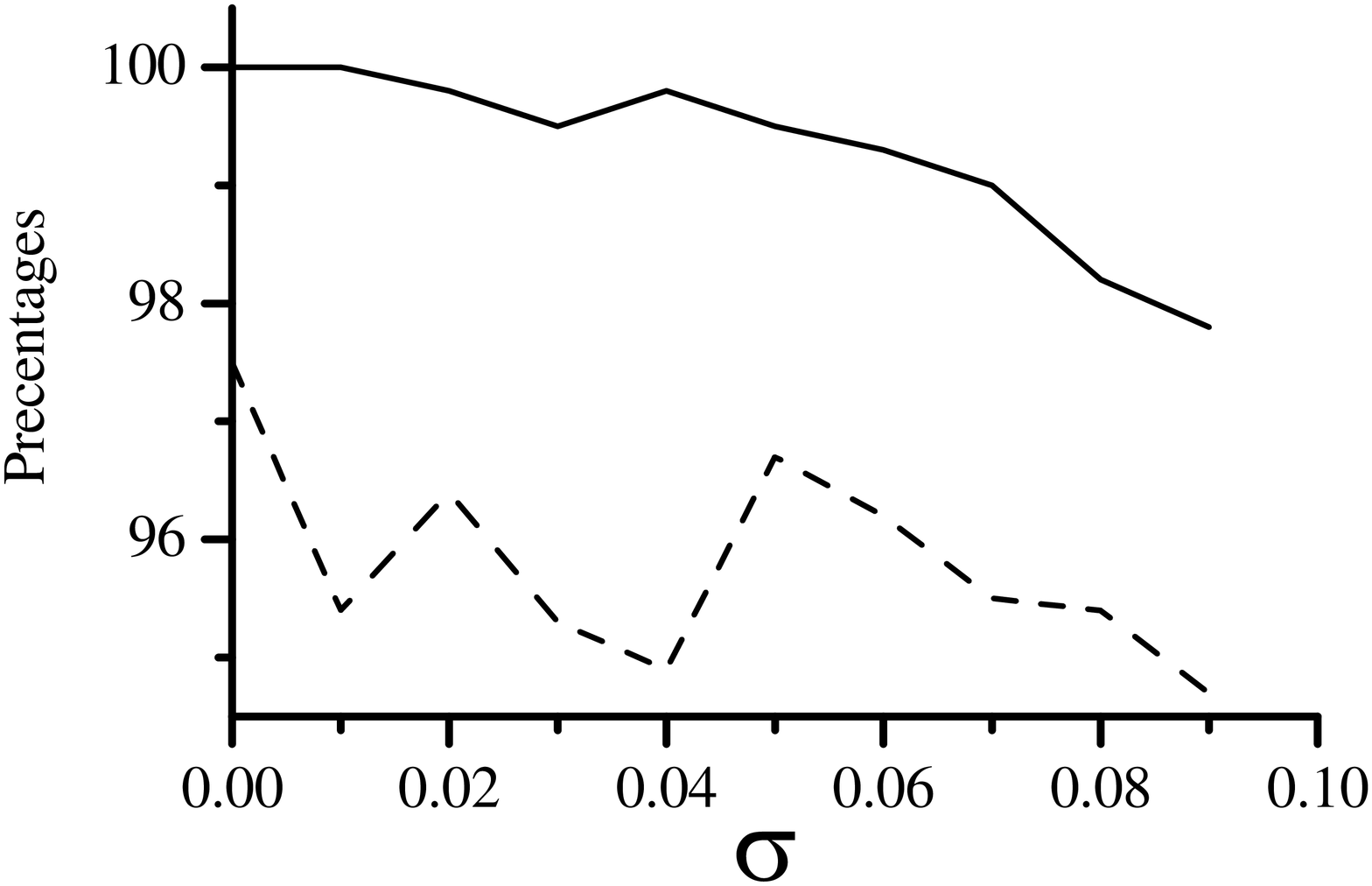}
\caption{\label{Noisy}Success percentage for the third set of simulations with
$F_0=1$ GHz and $F_{\textrm{nyq}}=20$ GHz as a function of standard deviation
$\sigma$ of the added noise. The figure shows the band-detection
percentage (solid curve) and reconstruction percentages (dashed
curve) in the case that both the maximum number of originating and
assumed positive bands equals 4.}
\end{figure}

In the this set of simulations, we reconstructed  signals with
different noise levels added. We chose $\xi=6$ MHz. The threshold
was chosen to be $T=2\max_i(\sigma^i)$. Accordingly, the parameter
$\rho$ in Eq.~\ref{W_smooth} was chosen to be $\rho=
\max_i(\sigma^i)$. The other parameters used in the simulation were
$a=2$ and $b=16$ MHz. Because the signals were not ideal, an exact
reconstruction was not possible and the definitions of an accurate
band detection and accurate reconstruction needed to be changed. A
band detection was deemed accurate if the originating bands
approximately matched the reconstructed bands. A signal
reconstruction was deemed accurate if the signal's originating bands
were detected accurately and if each reconstructed band
$X_{\mathcal{U}}(f)$ satisfied
\begin{equation}
\int_{B_m}|X_{\mathcal{U}}(f)-X(f)|<\max_i(\sigma^i)B_m.
\end{equation}
Here $X(f)$ is the noiseless signal and the integration is performed
over only the detected band. In a correct reconstruction, it is
expected that the average reconstruction error is lower than the
standard deviation of the noise in the noisiest channel, i.e. the
channel at the lowest sampling rate. We chose the same sampling
rates as those chosen in the second set of simulations. For these
rates: $\max_i(\sigma^i) =3.3\sigma$.

The detection percentages and reconstruction percentages are shown
in Fig.~\ref{Noisy}. The figure clearly shows that high percentages
are obtained even in the case of low signal to noise ratio. We
repeated this last\\[10cm] set of simulations using Gaussian signals instead of
the signals of Eq.~\ref{signal_cos}. We found that results are not
sensitive to the specific choice of signal type.

\section{Conclusion}

Typical undersampling schemes are PNS schemes. In such schemes
samples are taken from several channels at the same low rate. These
schemes have many drawbacks. In this paper we propose a new scheme
for reconstructing multi-band signals under the constraint of a
small number of sampling channels.  We have developed an MRS scheme;
a scheme in which each channel samples at a different rate. We have
demonstrated that sampling with our MRS scheme can overcome many of
the difficulties inherent in PNS schemes and can effectively
reconstruct signals from undersampled data. For a typical sparse
multi-band signal, our MRS scheme has the advantage over PNS schemes
because in almost all cases, the signal spectrum is unaliased in at
least one of the channels. This is in contrast to PNS schemes. With
PNS schemes an alias in one channel is equivalent to an alias in all
channels.

Our MRS scheme uses a smaller number of sampling channels than do
PNS schemes. We also choose to sample at a higher sampling rate than
PNS schemes use in attaining the theoretical minimum overall sampling rate
required for a perfect reconstruction. The use of higher rates has
an inherent advantage in that it increases the sampled signal to
noise ratio. Our MRS scheme also does not require the solving of
poorly conditioned linear equations that PNS schemes must solve.
This eliminates one source of lack of robustness of PNS schemes. Our
simulations indicate that our MRS scheme, using a small number of
sampling channels (3 in our simulations) is robust both to
different signal types and to relatively noisy signals.

Our reconstruction scheme does not require the synchronization of
different sampling channels. This significantly reduces the
complexity of the sampling hardware. Moreover, asynchronous sampling
does not require very low jitter between the sampling time at
different channels as is required in PNS schemes. Our reconstruction
scheme resolves aliasing in almost all cases but not all. In rare
cases, reconstruction of the originating signal fails owing to
aliasing. One of the methods to resolve aliasing is to synchronize
the sampling in all the channels. With such synchronization,
aliasing can be resolved by inverting a matrix similarly to as is
done in PNS schemes. However, such an approach requires both much
more complex hardware and a larger number of sampling channels that
sample with a very low jitter. Moreover, in case of signals that are
aliased simultaneously in all channels, the noise in the
reconstructed signal is expected to be much stronger than the noise
in the original signal.

Future work should focus on testing our algorithm's ability to
reconstruct experimental data. Optical systems for performing
experiments are currently in existence.

\section{Appendix}
In section 3.A.1 we have denoted the intervals over which the
indicator function \blue{$\mathcal{I}(f)=1$} by $U_1\ldots U_K$. In this
appendix we give a sufficient and necessary conditions under which
the spectral support of a signal coincides with a subset
$\mathcal{U}$ of $\{U_1,\ldots U_K\}$ and under which the function
$E_1(\mathcal{U})$ (Eq.~\ref{C}) is equal to zero. Although it
applies for more general cases, we assume that the function $X(f)$
is piecewise continuous.

The conditions are as follows:
\begin{enumerate}
  \item For each frequency $f_0$ which fulfills $\int_{f_0-\varepsilon}^{f_0+\varepsilon}
  |X(f)|^2 df >0$ for all $\varepsilon>0$, we obtain that\\
  $\int_{f_0-\varepsilon}^{f_0+\varepsilon} |X^i(f)|^2 df >0$ for all
  $\varepsilon>0$ and $1\leq i \leq P$.
  \item For each originating band with support $[a,b]$, there exists an
  interval $[a-\varepsilon,a+\varepsilon]$, ($\varepsilon\neq0)$ whose down-converted band does not overlap
  any other down-converted band in at least one of the sampled signals.
  Similarly, for each  originating band with  support $[a,b]$, there exists an
  interval $[b-\varepsilon,b+\varepsilon]$, whose down-converted band does not overlap
  any  other down-converted band in at least one of the sampled
  signals.
 \end{enumerate}
Condition 1 assures that originating bands are contained within
$\cup_{i=1}^{K} U_i$. Condition 2 guarantees that the originating
coincide exactly with a subset of $\mathcal{P}\{U\}$. It is obvious
that when the conditions are satisfied, $E_1(\mathcal{U})=0$.

The first condition excludes cases in which the down-converted bands
cancel each other's energy over a certain interval due to
destructive interference. When the condition is fulfilled, for each
frequency $f_0$ within the originating bands, we obtain
\blue{$\mathcal{I}(f_0)=1$}. Thus, each originating band $[a,b]$ is
contained within one of the intervals that make up the support of
\blue{$\mathcal{I}(f)$}. Mathematically, for each $[a,b]$, there exist
$U_k$, such that $[a,b] \subseteq U_k$.

The second conditions assures us that for each originating band
$[a,b]$, the intervals $[a-\varepsilon,a]$ and $[b,b+\varepsilon]$
are not contained within any of the $U_k$ for all values of
$\varepsilon$. Consequentially, if $[a,b] \subseteq U_k$, then $[a,b]=
U_k$. When the two conditions are fulfilled, we obtain that there
exist a set of intervals $\mathcal{U}$, which matches the
originating bands, and for which $E_1(\mathcal{U})=0$


\end{document}